\documentstyle [12pt,epsf] {article}

\newcommand{\nwc}{\newcommand}
\nwc{\cl}  {$\clubsuit$}

\nwc{\hyp} {\hyphenation}
\nwc{\be}  {\begin{equation}}
\nwc{\ee}  {\end{equation}}
\nwc{\ba}  {\begin{array}}
\nwc{\ea}  {\end{array}}
\nwc{\bdm} {\begin{displaymath}}
\nwc{\edm} {\end{displaymath}}
\nwc{\bea} {\be\ba{rcl}}
\nwc{\eea} {\ea\ee}
\nwc{\ben} {\begin{eqnarray}}
\nwc{\een} {\end{eqnarray}}
\nwc{\bda} {\bdm\ba{lcl}}
\nwc{\eda} {\ea\edm}
\nwc{\bc}  {\begin{center}}
\nwc{\ec}  {\end{center}}
\nwc{\ds}  {\displaystyle}
\nwc{\bmat}{\left(\ba}
\nwc{\emat}{\ea\right)}
\nwc{\non} {\nonumber}
\nwc{\bib} {\bibitem}
\nwc{\lra} {\longrightarrow}
\nwc{\Llra}{\Longleftrightarrow}
\nwc{\ra}  {\rightarrow}
\nwc{\Ra}  {\Rightarrow}
\nwc{\lmt} {\longmapsto}
\nwc{\prl} {\partial}
\nwc{\iy}  {\infty}
\nwc{\ol}  {\overline}
\nwc{\hm}  {\hspace{3mm}}
\nwc{\lf}  {\left}
\nwc{\ri}  {\right}
\nwc{\lm}  {\limits}
\nwc{\lb}  {\lbrack}
\nwc{\rb}  {\rbrack}
\nwc{\ov}  {\over}
\nwc{\pri}  {\prime}
\nwc{\nnn} {\nonumber \vspace{.2cm} \\ }
\nwc{\Sc}  {{\cal S}}
\nwc{\Lc}  {{\cal L}}
\nwc{\Rc}  {{\cal R}}
\nwc{\Dc}  {{\cal D}}
\nwc{\Oc}  {{\cal O}}
\nwc{\Cc}  {{\cal C}}
\nwc{\Pc}  {{\cal P}}
\nwc{\Mc}  {{\cal M}}
\nwc{\Ec}  {{\cal E}}
\nwc{\Fc}  {{\cal F}}
\nwc{\Hc}  {{\cal H}}
\nwc{\Kc}  {{\cal K}}
\nwc{\Xc}  {{\cal X}}
\nwc{\Gc}  {{\cal G}}
\nwc{\Zc}  {{\cal Z}}
\nwc{\Nc}  {{\cal N}}
\nwc{\fca} {{\cal f}}
\nwc{\xc}  {{\cal x}}
\nwc{\Ac}  {{\cal A}}
\nwc{\Bc}  {{\cal B}}
\nwc{\Uc}  {{\cal U}}
\nwc{\Vc}  {{\cal V}}
\nwc{\Th} {\Theta}
\nwc{\th} {\theta}
\nwc{\vth} {\vartheta}
\nwc{\eps}{\epsilon}
\nwc{\si} {\sigma}
\nwc{\Gm} {\Gamma}
\nwc{\gm} {\gamma}
\nwc{\bt} {\beta}
\nwc{\La} {\Lambda}
\nwc{\la} {\lambda}
\nwc{\om} {\omega}
\nwc{\Om} {\Omega}
\nwc{\dt} {\delta}
\nwc{\Si} {\Sigma}
\nwc{\Dt} {\Delta}
\nwc{\al} {\alpha}
\nwc{\vp} {\varphi}
\nwc{\kp} {\kappa}
\def\Tr{\mathop{\rm Tr}}

\def\gtap{\raisebox{-.4ex}{\rlap{$\sim$}} \raisebox{.4ex}{$>$}}
\nwc{\Id}  {{\bf 1}}
\nwc{\diag} {{\rm diag}}
\nwc{\inv}  {{\rm inv}}
\nwc{\mod}  {{\rm mod}}
\nwc{\hal} {\frac{1}{2}}
\nwc{\tpi}  {2\pi i}

\def\mpla#1{Mod.\ Phys.\ Lett.\ {\bf A#1}}
\def\npb#1{Nucl.\ Phys.\ {\bf B#1}}

\def\plb#1{Phys.\ Lett.\ {\bf B#1}}

\def\pra#1{Phys.\ Rev.\ {\bf A#1 }}
\def\prb#1{Phys.\ Rev.\ {\bf B#1 }}

\def \gta {\mathrel{\vcenter
     {\hbox{$>$}\nointerlineskip\hbox{$\sim$}}}}
\newsavebox{\nnin} \sbox{\nnin}{$\hspace{1mm}\in\kern -.8em /
                   \hspace{1mm}$}

\newcommand{\sub}{\subset}
\newsavebox{\nnsub} \sbox{\nnsub}{$\hspace{1mm}\sub\kern -.9em /
            \hspace{1mm}$}

\def\KK{{\rm I\kern -.2em  K}}
\def\NN{{\rm I\kern -.16em N}}
\def\RR{{\rm I\kern -.2em  R}}
\def\ZZ{Z \kern -.43em Z}
\def\QQ{{\rm \kern .25em
             \vrule height1.4ex depth-.12ex width.06em\kern-.31em Q}}
\def\CC{{\rm \kern .25em
             \vrule height1.4ex depth-.12ex width.06em\kern-.31em C}}
\def\ZZZ{Z\kern -0.31em Z}

\nwc{\olnu}  {\ol{\nu}}
\nwc{\olla}  {\ol{\la}}
\nwc{\olm}   {\ol{m}}
\nwc{\olmu}  {\ol{\mu}}
\nwc{\olh}   {\ol{h}}
\nwc{\olpsi} {\ol{\psi}}
\nwc{\olsi}  {\ol{\sigma}}
\nwc{\olgm}  {\ol{\gm}}
\nwc{\prlt}  {\frac{\prl}{\prl t}}
\nwc{\ttau}  {\tilde{\tau}}
\nwc{\trho}  {\tilde{\rho}}
\nwc{\tP}    {\tilde{P}}
\nwc{\tU}    {\tilde{U}}
\nwc{\teps}  {\tilde{\eps}}
\nwc{\tla}   {\tilde{\la}}
\nwc{\tit}    {\tilde{t}}
\nwc{\iddq}  {\int\frac{d^dq}{(2\pi)^d}}
\nwc{\prpr}  {\prime\prime}
\nwc{\rN}    {\left(\frac{\rho}{N}\right)}
\nwc{\rNt}    {\left(\frac{\rho}{N}\right)^{\frac{N-2}{2}}}
\nwc{\rnN}   {\left(\frac{\rho_0}{N}\right)}
\nwc{\rnNt}    {\left(\frac{\rho_0}{N}\right)^{\frac{N-2}{2}}}
\nwc{\rnNf}    {\left(\frac{\rho_0}{N}\right)^{\frac{N-4}{2}}}
\nwc{\rNs}    {\left(\frac{\rho_0}{N}\right)^{\frac{N-6}{2}}}
\nwc{\kNt}    {\left(\frac{\kappa}{N}\right)^{\frac{N-2}{2}}}
\nwc{\kNf}    {\left(\frac{\kappa}{N}\right)^{\frac{N-4}{2}}}
\nwc{\kNs}    {\left(\frac{\kappa}{N}\right)^{\frac{N-6}{2}}}

\newcounter{app}
\def\app{\par
 \addtocounter{app}{1}
 \def\thesection{\Alph{app}}
 \def\ksection{\Alph{app}}}

\def\appendix#1{\app\sect{#1}}

\begin{document}

\baselineskip14pt

\begin{center} 

FIELD THEORY NEAR THE CRITICAL TEMPERATURE\footnote[1]{Talk
given at the International School of Subnuclear Physics,
34th Course:
Effective Theories and Fundamental Interactions, Erice-Sicily,
3-12 July 1996.}\\

\vspace{5ex}

J\"URGEN BERGES\\
{\em Institut f\"ur Theoretische Physik,
Universit\"at Heidelberg}\\
{\em Philosophenweg 16, 69120 Heidelberg, Germany}

\end{center}

\vspace{0.5ex}

\begin{quotation}

\centerline{ABSTRACT}

\baselineskip12pt \footnotesize 

Field theory at nonvanishing temperature beyond perturbation theory
is discussed for the $N$-component $O(N)$-symmetric scalar
theory. We compute the effective potential directly in three
dimensions using
an exact evolution equation for an
effective action with an infrared cutoff.
A suitable truncation is solved numerically. 
We obtain a detailed quantitative picture of the scaling 
form of the equation
of state in the vicinity of the critical temperature of a second 
order phase transition.

\end{quotation}

\baselineskip14pt

\noindent
{\bf 1. Introduction}

The most prominent application of field theory at nonzero
temperature in particle physics concerns the symmetry
restoration in spontaneously broken gauge theories at high
temperature. In the standard model of electroweak interactions
gauge bosons and fermions acquire a mass due to a nonvanishing
vacuum expectation value of the Higgs doublet. According to 
arguments first given by Kirzhnitz and Linde$^1$
this expectation value vanishes at sufficiently high temperature
and the electroweak symmetry is restored. This phase 
transition\footnote[7]{For large Higgs boson masses 
around $m_H \gta 80$ GeV
there might be no phase transition
but rather an analytical crossover.$^{2,3,4}$} is 
expected to have occured in the early universe
which may have important consequences like 
the possible creation of the excess of
matter compared to antimatter (baryon asymmetry)$^5$. Another
challenge is
the study of QCD at nonvanishing temperature which is also relevant
for heavy ion collisions in the laboratory. At temperatures
around a few hundred MeV one expects dramatic changes in the 
structure and symmetries of hadronic matter$^6$. Of course, there are
other fields of application of nonzero temperature field theory as for 
instance condensed matter physics.

The prototype for investigations concerning the symmetry
restoration at high temperature is the $N$-component
scalar field theory with $O(N)$-symmetry.
For $N=4$ it describes the scalar
sector of the electroweak standard model in the limit of
vanishing gauge and Yukawa couplings. It is
also used as an effective model for the chiral
phase transition in QCD in the limit of two quark flavors$^7$.
In condensed matter physics $N=3$ corresponds to the well
known Heisenberg model used to describe the ferromagnetic
phase transition. There are other applications like the
helium superfluid transition ($N=2$), liquid-vapor transition
($N=1$) or statistical properties of long polymer chains
($N=0$).$^8$ 

The equation of state (EOS)
for a magnetic system is specified by the
free energy density per volume (here denoted by $U$) as a function
of arbitrary magnetization $\phi$ and temperature $T$.
All thermodynamic quantities can be derived from the function
$U(\phi,T)$. For example, the response of the system to
a homogeneous magnetic field $H$ follows from 
$\partial U/ \partial \phi = H$. This permits the computation
of $\phi$ for arbitrary $H$ and $T$.
There is a close 
analogy to quantum field theory at nonvanishing temperature.
Here $U$ corresponds to the temperature dependent effective
potential as a function of a scalar field $\phi$.
For instance, in the $O(4)$-symmetric model for the chiral
phase transition in two flavor QCD the meson field $\phi$
has four components. In this picture, quark masses are associated
with the source $H \sim m_q$ 
and one is interested in the behavior during the phase 
transition (or crossover) for $H \not= 0$.
The temperature and source
dependent meson masses and zero momentum interactions 
are given by derivatives of $U$.$^7$

The applicability of the $O(N)$-symmetric
scalar model to a wide class of very different
physical systems in the vicinity of the critical temperature
$T_{cr}$ at which the phase transition occurs
is a manifestation of universality of critical phenomena.
There exists a universal scaling form of the EOS
in the vicinity of a second order phase transition
which is not accessible to 
standard perturbation theory. 
The quantitative description of the scaling
form of the EOS will be the main topic of this talk.
To compute the effective potential $U$ which encodes
the EOS we employ a nonperturbative
method. It is based on an exact evolution
equation for an effective action $\Gm_k$
with an infrared cutoff $\sim k$. 
From $\Gm_k$ the standard effective action $\Gm$, i.e.\
the generating functional of $1PI$ Green functions, is
recovered when the additional infrared cutoff $\sim k$ 
is removed. Though I will
refrain from going into technical details, an introduction
to the method and 
some results for the EOS$^9$ will be presented.\\

\noindent
{\bf 2. Effective three dimensional behavior of high temperature
field theory} 

Quantum field theory
at nonzero temperature $T$ can be formulated in terms of an
Euclidean functional integral where the 'time' dimension 
is compactified
on a torus with radius $T^{-1}$.$^{10}$ If the characteristic length
scale of the considered physical problem is much larger than the
inverse temperature the compactified dimension cannot be
resolved ('dimensional reduction'$^{11}$).
One therefore observes an effective three dimensional
behavior of the high temperature quantum field theory. If the
phase transition is second order the correlation length becomes 
infinite and the effective three dimensional quantum
field theory is dominated by classical statistics.
In particular, the critical exponents which describe the
singular behavior of various quantities near the phase
transition are those of the corresponding classical statistical
system.$^{12}$ 
Analogous considerations are valid for sufficiently weak
first order phase transitions.

The calculation of the effective potential $U(\phi,T)$ in the
vicinity of the critical
temperature of a second order phase transition is an 
old problem. One can prove 
through a general
renormalization group analysis$^{13}$ the Widom
scaling form$^{14}$ of the EOS  
\be
H = \phi^{\delta} f
\left((T-T_{cr})/\phi^{1/\beta}\right) \label{wid}.
\ee
Only the 
limiting cases $\phi \rightarrow 0$ and $\phi \rightarrow \infty$
are quantitatively well described by 
critical exponents and amplitudes.
The critical exponents $\beta$ and $\delta$ have been
computed with high accuracy$^{8,15}$ but the scaling function
$f$ is more difficult to access.
A particular difficulty for a direct computation in three 
dimensions arises from the existence of massless Goldstone modes
in the phase with spontaneous symmetry breaking for models with continuous
symmetry ($N>1$).
They introduce severe infrared divergencies$^{16}$ within perturbative 
expansions. 
A related example for the breakdown of perturbation theory
is the symmetric phase of the electroweak theory where the gauge
bosons are massless. In addition, within a perturbative
description one is restricted to small expansion parameters.
For instance,
in the three dimensional $O(N)$-symmetric model no small
coupling characterizes the interactions at the phase transition.
One therefore has to employ nonperturbative methods which resolve
the infrared problem.

A standard method to cope with infrared singularities is the
'$\eps$-expansion' which is a double series expansion in powers 
of the coupling constant $g$ and $\eps = 4-d$. 
It has been used to compute the scaling function $f$
in second order in $\eps$ (third order for $N=1$)$^8$.
To study the 
three dimensional theory one sets the expansion parameter $\eps=1$.
The validity of this expansion is not clear a priori since the
expansion parameter is not small.
A nonperturbative computational method is provided by 
lattice Monte-Carlo simulations$^{19}$.
Nevertheless second order
or weak first order phase transitions are very demanding 
due to a finite lattice size and an infinite or large correlation 
length. In particular, it is notoriously difficult to 
distinguish by these methods between a weak first order and
a second order nature of the transition. The nonperturbative method
which allows here an unambiguous answer and which we will consider
in the following relies
on the use of an 'exact renormalization group
equation' (ERGE). ERGEs have been formulated in many different 
but formally equivalent ways$^{13,17}$. 
Though these equations are exact there is practically no chance
to solve them without approximations. The main challenge
is to find a suitable approximation scheme that describes all the
relevant physics of the considered problem. It is this point where
the particular formulation of an ERGE becomes important.
Here we will use the effective average 
action method$^{18}$ which allows nonperturbative 
calculations in a feasible 
way. Before turning to the method in section 4
we consider some of its results
for the scaling EOS for $O(N)$-symmetric models.$^{9}$ They
have been obtained from an approximate numerical solution
of the effective potential $U$ directly in three dimensions 
(cf.\ sect.\ 4).\\

\noindent
{\bf 3. Scaling equation of state}

Eq.\ (\ref{wid}) establishes the scaling properties of the
EOS. They have been explicitly verified by our numerical 
solution$^{9}$. In particular, the function $f$ depends
on only one scaling variable $x$. In our conventions
$x=-\dt \kp_{\La}/\phi^{1/\beta}$ where $\dt \kp_{\La}$ is
proportional to the deviation from the critical temperature
$\dt \kp_{\La} = A(T) (T_{cr}- T)$ with $A(T_{cr})~>~ 0$
and $\phi=\sqrt{\phi_a\phi^a}$ ($a=1,\ldots,N$). 
If one plots the logarithm of $f$ as a function
of the logarithm of its argument $x$ one can easily consider 
some limits and compare with known values of critical 
exponents and amplitudes. Figs.\ 1 and 2 show our results for
log$(f)$ and log$(df/dx)$ as a function of log$|x|$ for
$N=1$ and $N=3$. Here $x>0$
corresponds to temperatures above the critical temperature
(symmetric phase) and $x < 0$ accordingly to $T < T_{cr}$
(phase with spontaneous symmetry breaking).
\begin{figure}
\leavevmode
\centering  
\epsfxsize=4in
\epsffile{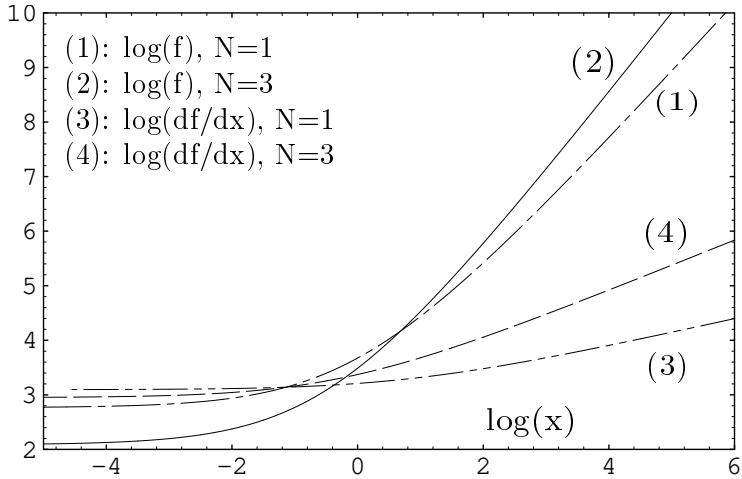}                                   
\caption{\footnotesize Logarithmic plot of the scaling function 
$f$ in the symmetric phase.}
\end{figure}

As an example we consider the limit $x \to \infty$. One 
observes that log$(f)$ becomes a linear function of log$(x)$  
with constant slope $\gamma$.
In this limit the universal function takes the form
\be
\lim\limits_{x \to \infty} f(x) = (C^+)^{-1} x^{\gamma} .
\ee     
The amplitude $C^+$ and the critical exponent $\gamma$
characterize the behavior of the 'unrenormalized' squared mass
or inverse susceptibility
\be 
\bar{m}^2=\chi^{-1}=\lim_{\phi \to 0}
\left(\frac{\prl^2 U}{\prl \phi^2}=
(C^+)^{-1}|\dt \kp_{\La}|^{\gamma}\phi^{\delta-1-\gamma/\beta}
\right).
\ee   
We have verified the scaling relation $\gamma/\beta=\delta-1$
that connects $\gamma$ with the exponents $\beta$ and $\delta$
appearing in the Widom scaling form (\ref{wid}). For the
singular behavior of the susceptibility for $\dt \kp_{\La} \to 0$ in the 
symmetric phase one finds $\chi = C^+ |\dt \kp_{\La}|^{-\gamma}$
with $\gamma=1.258 (1.465)$ for $N=1(3)$. In another limit, 
$x \to 0$, the scaling function becomes a constant, $f(0)=D$, and
according to eq.\ (\ref{wid}) one finds
$H = D \phi^{\dt}$ ($\dt=4.75(4.78)$ for $N=1(3)$).

The spontaneously broken phase is characterized by a nonzero
value $\phi_0$ of the minimum of the effective potential 
$U$ with
$H=(\prl U/\prl \phi) (\phi_0)=0$. The appearance
of spontaneous symmetry breaking below $T_{cr}$ implies
that $f(x)$ has a zero $x=-B^{-1/\beta}$ and one observes
a singularity of the logarithmic plot in fig.\ 2. In
particular, according to eq.\ (\ref{wid}) the minimum
behaves as $\phi_0 = B (\dt \kp_{\La})^{\beta}$ 
($\beta=0.336(0.388)$ for $N=1(3)$). 
\begin{figure}
\leavevmode
\centering  
\epsfxsize=4in
\epsffile{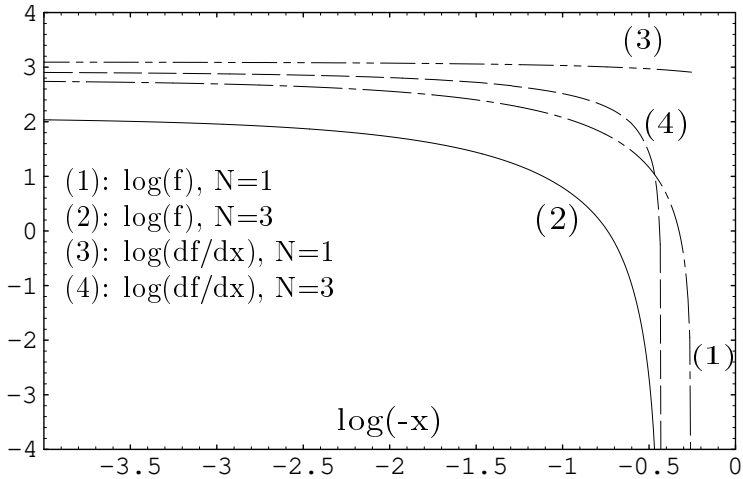}                                   
\caption{\footnotesize Logarithmic plot of the scaling function 
$f$ in the spontaneously broken phase.}
\end{figure}
The critical exponents like $\gamma$, $\beta$ and $\delta$ 
are universal whereas
the scaling function $f$ is universal up to arbitrary 
normalizations of $x$ and $f(x)$.
There are only two independent
scales in the vicinity of the transition point which can be related
to the deviation from the critical temperature and to the external
source. All models in the
same universality class can be related by a multiplicative
rescaling of $\phi$ and $\dt \kp_{\La}$ or $(T_{cr}-T)$.
Accordingly there are only two independent amplitudes
and exponents respectively.  

Apart from the given examples we have extracted other
quantities which characterize the asymptotic behavior
for the sake of comparison.
The critical exponents and amplitude ratios
we have calculated typically
deviate by a few per cent from the values obtained by other 
methods$^{8,15}$. We expect the error to be related to the size of the
anomalous dimension $\eta \simeq 4 \%$. 
For recent lattice results for the EOS and a comparison
see ref.\ 19.
A more detailed discussion
of the EOS, including semi-analytical expressions for 
the scaling function $f(x)$
for $N=1$ and $N=3$ and an alternative parametrization of the 
EOS in terms of renormalized quantities,
can be found in ref.\ 9.\\

\noindent
{\bf 4. Effective average action and exact evolution equation}

In the following we turn to the method we have used to derive
the above results.
It relies on 
the effective average action$^{18}$ $\Gm_k$. 
For a theory described by a classical action $S$, 
the effective average action results from the 
integration of fluctuations with characteristic momenta larger
than a given infrared cutoff $\sim k$. 
To be explicit 
we consider a $k$-dependent
generating functional with $N$ real scalar fields 
$\chi_a$,
\be
W_k[J]=\ln \int D \chi \exp\left(-S[\chi]-\Dt_k S[\chi]+
\int d^dx J_a(x)\chi^a(x) \right)
\ee
where the k-dependence arises from the additional term
\be
\Dt_k S[\chi]=\hal \int 
\frac{d^dq}{(2\pi)^d} R_k(q)\chi_a(-q)\chi^a(q).
\ee
Without this term $W_k$ becomes the usual generating functional
for the connected Green functions. Here the 
infrared cutoff function $R_k$ is required to vanish
for $k \to 0$ and to diverge for $k \to \infty$
and fixed $q^2$.
This can be achieved, for example, by the choice
$R_k(q)=Z_k q^2 e^{-q^2/k^2}(1- e^{-q^2/k^2})^{-1}$
where $Z_k$ denotes an appropriate wave function renormalization
constant which will be defined below.
For fluctuations with small momenta
$q^2\ll k^2$ the cutoff $R_k\simeq Z_k k^2$ acts like an
additional mass term and prevents their propagation. 
For $q^2\gg k^2$ the infrared cutoff vanishes 
such that
the functional integration of the high momentum modes
is not disturbed. 
The expectation value of $\chi$ in the presence of $\Dt_k S[\chi]$
and $J$ reads $\phi^a \equiv <\chi^a> = \dt W_k[J]/\dt J_a$.
We define the effective average action as
\be
\Gm_k[\phi]=-W_k[J]+\int d^dx J_a(x)\phi^a(x)-\Dt_k S[\phi]. 
\ee
As a consequence, as the scale $k$ is lowered
$\Gm_k$ interpolates
from the classical action $S=\lim_{k \to \infty}\Gm_k$ to the 
standard effective action $\Gm=\lim_{k \to 0}\Gm_k$, i.e.\
the generating functional of $1PI$ Green functions.$^{18}$ 
Lowering $k$ results in 
a successive inclusion of fluctuations with momenta
$q^2 \gtap k^2$ and therefore permits to explore the theory on 
larger and larger length scales. One can view 
$\Gm_k$ as the effective action for averages of fields
over a volume of size $\sim k^{-d}$ and the approach is similar
in spirit to the block spin action$^{20,13}$ on the lattice. 
The interpolation property of $\Gm_k$ can be used to 'start' 
at some high momentum
scale $\La$ where $\Gm_{\La}$ can be taken as the classical or
short distance action and to solve the theory by following
$\Gm_k$ to $k \to 0$.
The scale dependence of $\Gm_k$ can be
described by an {\em exact} evolution equation$^{18}$
\be
 \prlt \Gm_k [\phi] =
 \hal\Tr\left\{\left(
 \Gm_k^{(2)}[\phi]+R_k\right)^{-1}
 \frac{\prl R_k}{\prl t}\right\}
 \label{ERGE}
\ee
where $t=\ln(k/\La)$.
The evolution is described in terms of the exact inverse
propagator $(\Gm_k^{(2)})^a_b(q,q')=\dt^2 \Gm_k/\dt \phi_a(-q)
\dt \phi^b(q')$. The trace involves a momentum 
integration as well as a summation
over internal indices. 
The additional cutoff function $R_k$ with a form like
the one given above renders the momentum integration both 
infrared (IR) and ultraviolet 
(UV) finite. In particular, the direct implementation of
the additional mass term $R_k \simeq Z_k k^2$ for $q^2 \ll k^2$
into the inverse average propagator makes the formulation suitable
for dealing with theories which are plagued by infrared problems
in perturbation theory. 

Though the evolution equation for the effective average
action is exact it remains a complicated functional
differential equation. In practice one has to find
a truncation for $\Gm_k$ and obtains approximate solutions.
Our ansatz represents the lowest order in a systematic
derivative expansion of $\Gm_k$,
\be
 \Gm_k = \ds{\int d^d x\left\{ U_k(\rho)+
 Z_k \prl_{\mu} \phi_a \prl^{\mu} \phi^a
  \right\} \qquad ,\,\,  a=1,\ldots,N.}
 \label{Ansatz}
\ee
Here $\phi^a$ denotes the $N$-component real scalar field and
$\rho = \frac{1}{2} \phi^a \phi_a$. We keep for the 
potential term the most general $O(N)$-symmetric 
form $U_k(\rho)$ since $U(\rho)=
\lim_{k \to 0} U_k(\rho)$ encodes the equation of state.
The wavefunction renormalization
is approximated by one $k$-dependent parameter $Z_k$.
The first correction to this ansatz would include field 
dependent wave function
renormalizations $Z_k(\rho)$ plus functions not
specified in eq.\ (\ref{Ansatz}) which account for a
different index structure of invariants with two
derivatives for $N>1$.$^{12,18}$ The next level involves 
invariants with four derivatives and so on.
Concerning the equation of state
the truncation of the higher derivative terms 
typically generates an uncertainty of the order of the anomalous
dimension $\eta$.$^{9}$
For $N=1$ the weak $\rho$-dependence 
of $Z_k$ has been established
explicitly at the critical temperature$^{21}$.
We define $Z_k$ 
at the minimum $\rho_0$ of $U_k$ and
at vanishing momenta $q^2$, i.e.\
$Z_k=Z_k(\rho=\rho_0;q^2=0)$.
The $k$-dependence of this function is determined by the anomalous
dimension $\eta(k)=-\mbox{d}(\mbox{ln}  Z_k)/\mbox{d}t$. 

If the ansatz (\ref{Ansatz}) is inserted into the 
evolution equation
for the effective average action (\ref{ERGE}) one obtains flow equations  
for the effective average potential $U_k(\rho)$ and for the
wave function renormalization constant $Z_k$ (or equivalently the 
anomalous dimension $\eta$). 
These 
have to be integrated starting from some short distance scale 
$\La$ and one has to specify $U_{\La}$ and $Z_{\La}$
as initial conditions. 
The short distance potential is taken to be a quartic potential
\be
U_{\La}(\rho)=-\bar{\mu}_{\La}^2 \rho + \hal \bar{\la}_{\La}
\rho^2 
\label{uinitial}
\ee 
and $Z_{\La}=1$. 
For a study of the behavior in the vicinity of the phase transition,
it is convenient to work
with dimensionless renormalized fields 
$\trho = Z_k k^{2-d} \rho$ and we also switch to a
dimensionless potential $u_k = k^{-d} U_k$.
The evolution equation for $u_k(\trho)$ reads$^{12,18,21}$ 
(primes denote derivatives with respect to $\trho$ at fixed $t$)
\be
\ds{\frac{\prl u_k}{\prl t} = -d u_k + (d-2+\eta) \trho u_k^{\pri}
+ 2 v_d (N-1) l^d_0(u_k^{\pri};\eta) + 2 v_d
l^d_0(u_k^{\pri}+2 \trho u_k^{\pri\pri};\eta)} \label{pot}
\ee
where $v_d^{-1}=2^{d+1}\pi^{d/2}\Gm(d/2)$ with $v_3 = 1/8 \pi^2$.
The anomalous dimension $\eta$ is given in our truncation by$^{12,18}$ 
\be
\eta(k) =  \frac{16 v_d}{d} \kp (u^{\pri\pri}_k(\kp))^{2} 
m^d_{2,2}(2 \kp u^{\pri\pri}_k(\kp);\eta)
\ee 
with $\kp$ the location of the minimum of the potential, $u'_k(\kp)= 0$.
The functions $l^d_0$ and $m^d_{2,2}$
result from the momentum integration in the flow equation
(\ref{ERGE}) for $\Gm_k$. They
equal constants of order one for vanishing
arguments and decay fast for arguments much larger than 
one.$^{12,22}$
As a consequence these functions account for the decoupling of 
modes with masses much larger than the 
scale $k$.
For the
minimum of the dimensionless potential at $\kp \not = 0$ 
one easily recognizes in eq.\ (\ref{pot}) the contribution
due to the $(N-1)$ massless Goldstone bosons 
with dimensionless mass term $u'_k(\kp)= 0$ and the radial mode
with dimensionless mass term $2 \kp u_k^{\pri\pri}(\kp)$.

At a second order phase transition 
the correlation length diverges and accordingly
there is no mass scale present in the theory.
In particular, one expects a scaling behavior of the effective 
average potential.  
Exactly at a second order phase transition $u_k$ should be 
given by a $k$-independent (scaling) solution $\prl u_k/\prl t=0$.
The EOS involves the potential away from the critical temperature.
Its computation therefore requires the solution of the full
partial differential equation (\ref{pot}) for the dependence
of $u_k$ on the two variables $\trho$ and $k$. Suitable 
numerical methods for its solution are discussed in ref.\ 22.

In the remaining part of this section we consider the theory
in three dimensions $(d=3)$
with the classical potential (\ref{uinitial})
as initial condition. For given $\bar{\la}_{\La}/\La$
there is a critical value for the minimum 
$\kp_{\La}=\bar{\mu}_{\La}^2/(\bar{\la}_{\La}\La)=\kp_{cr}$ 
of the classical potential for which
the scaling solution is approached in the
limit $k \to 0$. 
In particular, the minimum $\kp$ and
the anomalous dimension $\eta$ 
take on constant (fixed point) values
$\kp(k)= \kp_{\star}$ and $\eta(k)=\eta_{\star}$. 
The order of the phase transition 
becomes apparent from the observation that the order
parameter $\rho_0 \sim \lim_{k \to 0}
k^{1+\eta_{\star}} \kp_{\star} = 0$
vanishes at the phase transition.

To demonstrate the behavior of the scale dependent
effective potential near the phase transition
we consider the shape of $u_k^{\pri}(\trho)$
for different values of the scale.
In fig.\ 3 the numerical solution of the function 
$u^{\pri}_k(\trho)$ is plotted 
for $d=3$ and $N=1$
and various values of $t= \ln(k/\La)$.$^{22}$ The evolution starts
at $k=\La$ ($t=0$) where no integration of fluctuations
has been performed.
We arbitrarily choose $\bar{\la}_\La/
\La=0.1$ and fine tune $\kp_\La \simeq \kp_{cr}$ 
so that the scaling solution is approached at later stages 
of the evolution. If $\kp_{\La}$ is interpreted as a function of 
temperature the small deviation $\dt \kp_{\La}=\kp_{\La}-\kp_{cr}$ is
proportional to the deviation from the critical temperature.
\begin{figure}
\leavevmode
\centering  
\epsfxsize=4in
\epsffile{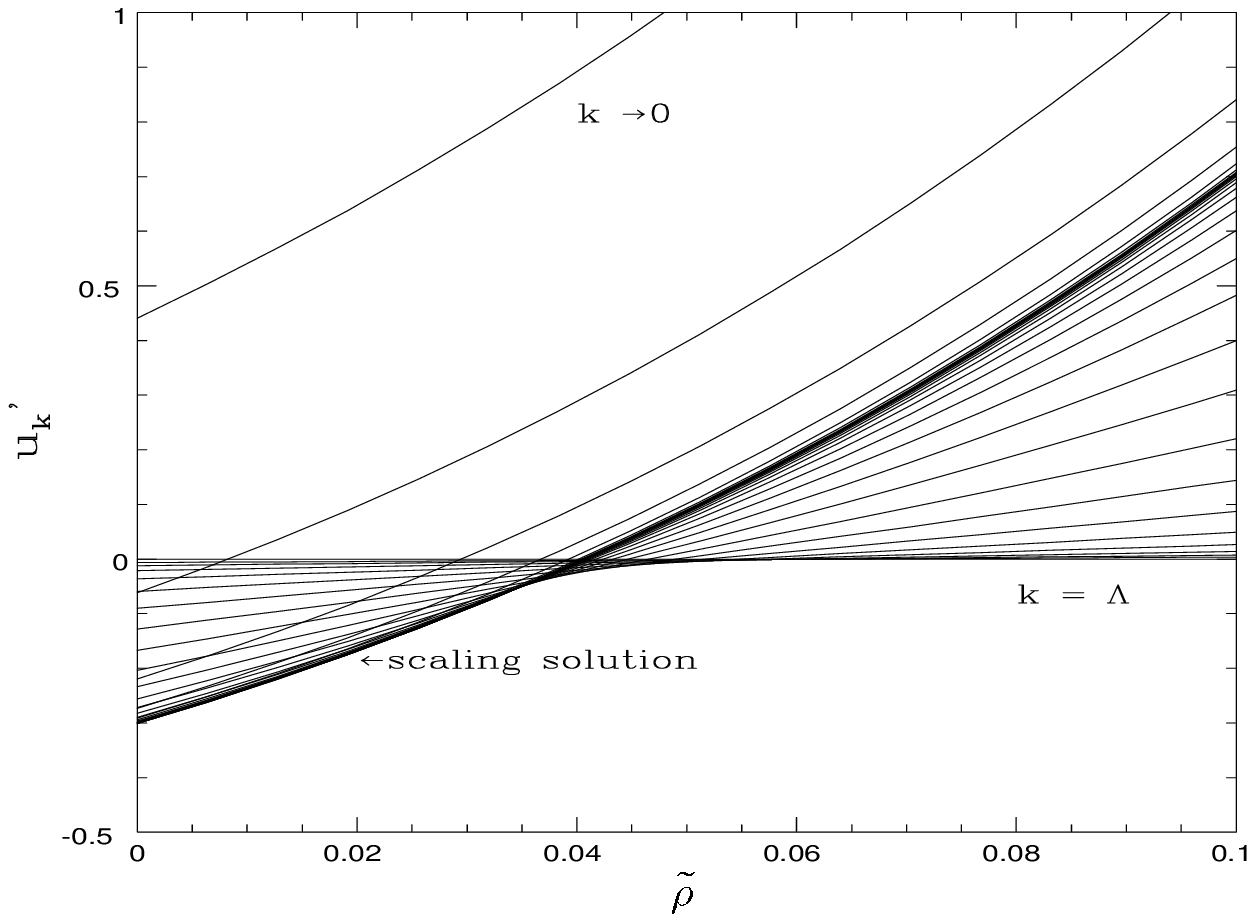}                                   
\caption{\footnotesize The evolution of $u^{\pri}_k(\trho)$
as k is lowered from $\La$ to zero.}
\end{figure}
For the example in fig.\ 3 a value slightly above the critical 
temperature is
used $(\dt \kp_{\La}<0)$. 
As $k$ is lowered $u'_k(\trho)$ deviates from the initial
linear shape of $u'_{\La}(\trho)$. 
By staying near the scaling solution for 
several orders of magnitude in $k$, the system looses memory
of the initial conditions at the short distance scale $\La$.
As a result, after $u'_k(\trho)$ has
evolved away from the scaling solution, its
shape is independent of the choice of $\bar{\la}_\La/\La$ for
the classical theory.
This property gives rise to the universal behavior
near second order phase transitions
discussed in section 3. Eventually, for
the final part of the evolution as 
$k \to 0$, the theory settles down either
in the symmetric phase $(\dt \kp_{\La}<0)$ 
as is demonstrated in fig.\ 3 with
$\kp =0$ and positive constant mass term 
$m^2=\lim_{k \to 0} k^2 u'_k(0) $, or 
$\kp$ grows in such a way that 
$\rho_0=\lim_{k \to 0} Z_k^{-1} k \kp$ 
approaches a constant value
indicating spontaneous symmetry breaking
$(\dt \kp_{\La}>0)$.
The scaling EOS obtains from 
$U^{\pri}(\rho)=\lim_{k \to 0} Z_k k^2 u'_k(\trho)$
with $H/\phi^{\dt}=U^{\pri}\phi^{\dt - 1}=f$ as
a function of $x=-\dt \kp_{\La}/\phi^{1/\bt}$ 
($\phi=\sqrt{2 \rho}$). \\

\noindent
{\bf 5. Conclusions and outlook}

We have arrived at a detailed quantitative picture of the
equation of state in the vicinity of the critical 
temperature of a second order phase transition. The method
we have used is free of IR or UV divergencies and provides 
a practical tool for calculations not accessible to 
perturbation theory. It relies on the effective average 
action $\Gm_k$ which interpolates between the classical 
action $S$ ($k=\La$) and the effective action $\Gm$ 
($k=0$). Its scale 
dependence is described by an exact 
nonperturbative evolution equation.

Though not discussed in the present talk, the method
can be applied as well to a description of first order
phase transitions along the lines
presented above. In particular, 
the scaling equation of state for weak (fluctuation
induced) first order phase transitions in scalar matrix
models has been obtained.$^{23}$ Work in progress focuses on the 
equation of state for an 
effective quark-meson model$^{24}$
at nonzero temperature$^{25}$, relevant for the study of the chiral
phase transition in QCD.

The average action approach allows a well-defined 
description of a rich spectrum of critical phenomena like the
observation of critical exponents, tricritical points and
crossover behavior. 
Further applications at nonzero temperature in various
levels of truncations include studies of the crossover behavior  
in two-scalar theories$^{26}$ or the description of
the Kosterlitz-Thouless phase transition
in two space dimensions$^{27}$. 
Extensions to 
gauge theories$^{3,28,29}$ 
have been e.g.\ applied to the
abelian Higgs model and to
the electroweak phase transition at nonzero temperature$^{3,28}$.
As should be pointed out
the method is not restricted to a discussion of the universal
behavior of the theory which can be observed near or at
second order or sufficiently weak first order phase transitions.
It also permits the description of nonuniversal physics such
as the behavior of the theory  
away from the critical temperature
for the four dimensional theory at nonvanishing 
temperature$^{12}$.\\

\noindent
{\bf 6. Acknowledgment}

I am grateful to N.\ Tetradis and C.\ Wetterich for collaboration on
the scaling EOS and 
in addition to J.\ Adams, S.\ Bornholdt and F.\ Freire for
collaboration on numerics.
I would like to express my gratitude to the organizers of this school
for providing a most stimulating environment.\\

\pagebreak
  
\noindent
{\bf 7. References}
\begin{enumerate}

\parsep0pt
\itemsep-4pt

\item D.A.\ Kirzhnitz, JETP Lett.\ {\bf 15} (1972) 529;
D.A.\ Kirzhnitz and A.D.\ Linde, \plb{72} (1972) 471.

\item See M.E.\ Shaposhnikov, these proceedings,
hep-ph/9610247; K.\ Kajantie, M.\ Laine,
K.\ Rummukainen and M.E.\ Shaposhnikov, hep-ph/9605288.

\item M.\ Reuter and C.\ Wetterich, Nucl.\ Phys.\ {\bf B408}
(1993) 91, ibid.\ {\bf 417} (1994) 181, ibid.\ {\bf 427} (1994) 
291; B.\ Bergerhoff and C.\ Wetterich, \npb{440} (1995) 91;
N.\ Tetradis, hep-ph/9608272.

\item W.\ Buchm\"uller and O.\ Philipsen, \npb{443} (1995) 47.

\item V.A.\ Kuzmin, V.A.\ Rubakov and M.E.\ Shaposhnikov,
\plb{155} (1985) 36; M.E.\ Shaposhnikov, \npb{287} (1987) 757;
ibid.\ {\bf 299} (1988) 797.

\item See e.g.\ A.\ Smilga, hep-ph/9604367.

\item R.D.\ Pisarski and F.\ Wilczek, Phys.\ Rev.\ 
{\bf D29} (1984) 338;
F.\ Wilczek, Int.\ J.\ Mod.\ Phys.\ {\bf A7} (1992) 3911;
K.\ Rajagopal and F.\ Wilczek, \npb{399} (1993) 395. 
 
\item J.\ Zinn-Justin, {\it Quantum Field Theory
and Critical Phenomena} (Oxford University Press,
1993), ch.\ 27 and references therein.

\item J.\ Berges, N.\ Tetradis and C.\ Wetterich, 
Phys.\ Rev.\ Lett.\ {\bf 77} (1996) 873.

\item J.\ Kapusta, {\it Finite Temperature Field Theory} 
(Cambridge University Press, 1989).

\item The topic of dimensional reduction is discussed by
M.E.\ Shaposhnikov, these proceedings, hep-ph/9610247.

\item N.\ Tetradis and C.\ Wetterich, Nucl.\ Phys.\ {\bf B398} 
(1993) 659; ibid.\ {\bf 422} (1994) 541; 
Int.\ J.\ Mod.\ Phys.\ {\bf A9} (1994) 4029.

\item K.G.\ Wilson, Phys.\ Rev.\ {\bf B4} (1971) 3174; 3184;
K.G.\ Wilson and I.G.\ Kogut, Phys.\ Rep.\ {\bf 12} (1974) 75;
F.J.\ Wegner, in {\it Phase Transitions and Critical Phenomena}, 
vol.\ 6, eds.\ C.\ Domb and M.S.\ Greene, 
(Academic Press, 1976). 

\item B.\ Widom, J.\ Chem.\ Phys.\ {\bf 43} (1965) 3898. 

\item 
P.\ Butera and M.\ Comi, hep-lat/9505027; T.\ Reisz, 
\plb{360} (1995) 77.

\item A.D.\ Linde, \plb{96} (1980) 289;
D.\ Gross, R.D.\ Pisarski and L.\ Yaffe, Rev.\ Mod.\ Phys.\ 
{\bf 53} (1981) 43.

\item F.\ Wegner and A.\ Houghton, \pra8 (1973) 401;
S.\ Weinberg in {\it Critical phenomena for field theorists}, Erice
Subnucl.\ Phys.\ 1 (1976);
J.F.\ Nicoll and T.S.\ Chang, Phys.\ Lett.\ {\bf A62} (1977) 287;
J.\ Polchinski, Nucl.\ Phys.\ {\bf B231} (1984) 269;
A.\ Hasenfratz and P.\ Hasenfratz, Nucl.\ Phys.\  {\bf B270} 
(1986)~687. 

\item C.\ Wetterich, Nucl.\ Phys.\  {\bf B352} (1991) 529;
Z.\ Phys.\ {\bf C57} (1993) 451; ibid  {\bf 60} (1993) 461;
\plb{301} (1993) 90.
  
\item For recent lattice results see M.M.\ Tsypin, 
Phys.\ Rev.\ Lett.\ {\bf 73} (1994) 2015;
hep-lat/9601021; D.\ Toussaint, hep-lat/9607084.

\item  L.P.\ Kadanoff, Physica {\bf 2}
(1966) 263.

\item T.\ R.\ Morris, Phys.\ Lett.\  {\bf B329} (1994) 241.

\item J.\ Adams, J.\ Berges, S.\ Bornholdt, F.\ Freire, N.\ Tetradis 
and C.\ Wetterich, \mpla{10} (1995) 2367.

\item J.\ Berges and C.\ Wetterich, hep-th/9609019.

\item D.-U.\ Jungnickel and C.\ Wetterich,
Phys.\ Rev.\ {\bf D53} (1996) 5142.

\item J.\ Berges, D.-U.\ Jungnickel, B.-J.\ Schaefer and C.\
Wetterich, in preparation.

\item S.\ Bornholdt, P.\ B\"uttner, N.\ Tetradis and C.\ Wetterich,
cond-mat/9603129; S.\ Bornholdt, N.\ Tetradis and C.\ Wetterich,
Phys.\ Rev.\ {\bf D53} (1996) 4552.

\item M. Gr\"ater and C.\ Wetterich, Phys.\ Rev.\ Lett.\ {\bf 75}
(1995) 378.

\item M.\ Reuter and C.\ Wetterich, Nucl.\ Phys.\ {\bf B391}
(1993) 147;
B.\ Bergerhoff, D.\ Litim, S.\ Lola
and C.\ Wetterich, Int.\ J.\ Mod.\ Phys.\ {\bf A11} (1996) 4273;
B.\ Bergerhoff, F.\ Freire, D.\ Litim, S.\ Lola and C.\ Wetterich,
\prb{53} (1996) 5734.

\item
M.\ Bonini, M.\ D' Attanasio and G.\ Marchesini,
\npb{418} (1994) 81; {\em ibid.} {\bf 421} (1994) 429;
{\em ibid.} {\bf 437} (1995) 163; \plb{346} (1995) 87;\\
U.\ Ellwanger, \plb{335} (1994) 364; 
U.\ Ellwanger, M.\ Hirsch and A.\ Weber, hep-ph/9606468.

\end{enumerate}

\end{document}